# Practical Solutions For Format-Preserving Encryption

Mor Weiss
Technion
Haifa, Israel
Email: morw@cs.technion.ac.il

Boris Rozenberg
IBM
Haifa, ISrael
Email: borisr@il.ibm.com

Muhammad Barham
IBM
Haifa, Israel
Email: muhammad@il.ibm.com

*Abstract*—Format Preserving Encryption (FPE) schemes encrypt a plaintext into a ciphertext while preserving its format (e.g., a valid social-security number is encrypted into a valid social-security number), thus allowing encrypted data to be stored and used in the same manner as unencrypted data. Motivated by the always-increasing use of cloud-computing and memory delegation, which require preserving both plaintext format and privacy, several FPE schemes for general formats have been previously suggested. However, current solutions are both insecure and inefficient in practice. We propose an *efficient* FPE scheme with *optimal security*. Our scheme includes an efficient method of representing general (complex) formats, and provides efficient encryption and decryption algorithms that do not require an expensive set-up. During encryption, only *format*-specific properties are preserved, while all *message*-specific properties remain hidden, thus guaranteeing data privacy. As experimental results show that in many cases large formats domains cannot be encrypted efficiently, we extend our scheme to support large formats, by imposing a user-defined bound on the maximal format size, thus obtaining a flexible security-efficiency tradeoff and the best possible security (under the size limitation).

## I. Introduction

Encryption schemes are used to protect data privacy, e.g., when transmitted over insecure channels or stored on unreliable servers. However, standard encryption schemes (such as AES) can significantly alter the data format, causing disruptions both in storing and using the data. Indeed, when storing devices and applications are designed to operate on unencrypted data they may not be able to operate on encrypted data. Consequently, Format-Preserving Encryption (FPE) schemes, namely schemes which encrypt messages into ciphertexts with the same format, have emerged as a most useful tool in applied cryptography.

CURRENT SOLUTIONS. First studied in the context of integral domains (namely, when the message domain is $\mathcal{M} = \{0, 1, ..., m-1\}$ for some $m \in \mathbb{N}$) [4], later works [2] considered more general formats, and two general techniques were suggested for FPE design. First, the *cycle walking* strategy of Black and Rogaway [4] constructs an FPE for format $\mathcal{F}$ from *any* FPE for a format $\mathcal{F}'$ such that $\mathcal{F} \subseteq \mathcal{F}'$. The encryption algorithm for $\mathcal{F}$ repeatedly applies the encryption algorithm of $\mathcal{F}'$, until the ciphertext is in $\mathcal{F}$. (Decryption is repeated until reaching a valid string in $\mathcal{F}$.) For example, an FPE scheme for credit-card numbers can use cycle-walking on AES (which is an FPE for $\{0, 1\}^{128}$).

Second, the *Rank-then-Encipher* (RtE) method suggested by Bellare et al. [2] reduces the task of designing an FPE for format $\mathcal{F}$ to the task of designing an FPE for an integral domain. (In particular, the RtE framework allows one to apply the same encryption logic to all formats, thus eliminating the need to design specially-tailored encryption schemes for every format.) More specifically, a format $\mathcal{F}$ of size $N$ is arbitrarily ordered as $\mathcal{F} = \{s_0, ..., s_{N-1}\}$, and encryption (decryption) is based on an integer-FPE (i.e., for an integral domain), where a string $s \in \mathcal{F}$ is encrypted in three steps, called ranking, integer-encryption, and unranking. First, the index $i$ such that $s = s_i$ is found; then $i$ is encrypted into an index $j$, using the integer-FPE encryption algorithm; finally, the encryption of $s$ is the message $s_j$. (Decryption is performed in the same manner by replacing the integer-FPE encryption with the decryption algorithm.) If $\mathcal{F}$ has a deterministic finite automaton (DFA), then $\mathcal{F}$ has efficiently computable ranking and unranking algorithms [6]. We note that the scheme inherits its security from the integer-FPE, while ranking and unranking do not contribute to security. Efficiency of the scheme relies heavily on the efficiency of ranking and unranking. The combination of cycle-walking and RtE yield an FPE scheme for any "rankable" format, which raises the question of designing *efficient* ranking and unranking methods for general formats. We stress that though ranking can use translation-tables, such tables cannot be constructed efficiently, require expensive storage, and do not admit efficient searching algorithms. Moreover, designing a single encryption scheme for several formats raises the question of efficiently representing formats, since a representation of the format will be given as input to the encryption algorithm.

LIMITATIONS OF CURRENT SOLUTIONS. Several works suggested FPEs for specific formats, such as fixed-base, fixed-length vectors (i.e., $\{0, 1, ..., m\}^n$ for $n, m \in \mathbb{N}$) [3], [5]; and more practical message-domains such as social-security numbers [10], credit-card numbers, and dates [8], but these schemes were tailored for *specific* formats, and it is not clear whether, and how, they can be generalized. The ranking strategy suggested for more general formats (e.g., names, addresses, etc.) [13], [1] partitions the format into many sub-formats, where the messages in each sub-format share additional characteristics (e.g., length), and therefore raises both efficiency and security concerns.

Regarding security, the scheme maintains "cosmetic" characteristics of the message which are not part of the properties defining the format, thus allowing an attacker to deduce (from the ciphertext) many message-specific characteristics which do not follow from $m$ having format $\mathcal{F}$. As we show in Section III-B, this renders the scheme completely insecure in theory *and* practice. Regarding efficiency, the scheme of [13], [1] is inefficient in practice. First, they do not suggest a method of efficiently representing formats, and partitioning the format into sub-formats (which must be done before encryption) is too costly to be performed in practice, since it depends on the *number* of plaintexts, rather than their *lengths* as in non-format-preserving encryption schemes. Second, all formats (even when $|\mathcal{F}| \gg 2^{128}$, which is the case for many practical formats) are encrypted using the same methods. As these schemes rely on integer-FPEs that are inefficient when applied to large domains, encrypting and decrypting large formats is inefficient in practice. Thirdly, as the scheme admits no method of representing complex format properties, encryption can be inefficient even for medium-sized format due to cycle-walking, which repeatedly applies the "heavy" operations of integer-FPE encryption and decryption. Therefore, the average cycle length may be long, and more importantly, there is no worst-case bound on the *actual* cycle length. This motivates eliminating the use of cycle-walking.

OUR CONTRIBUTIONS. We address both issues by proposing a RtE-based FPE scheme for general formats that is both efficient and secure. We provide a framework for defining complex formats according to their unique properties, thus avoiding format-partition (*except* when the format is too large to be encrypted efficiently without partitioning). This is done by identifying several simple formats, and defining composition operations which are used to represent complex formats. In addition, we provide a general logic for *inclusively* ranking and unranking complex formats (namely, ranking all ingredients of the format together, instead of ranking every building-block separately). Our framework offers a method of efficient format representation, thus eliminating the need for cycle-walking, and improving the average and actual efficiency. Moreover, by preserving only *format* properties (without preserving any *message-specific* properties), we get an (optimally) secure general FPE scheme. Finally, by incorporating a user-controlled bound on format sizes, we give a practical solution for large format encryption. Our scheme was also filed for US patent [11].

CONCURRENT WORK. In a concurrent work, Luchaup et al. [9] developed libFTE - a unifying format-preserving and format-*transforming* encryption scheme (in format-transforming encryption, all ciphertexts are guaranteed to have format $\mathcal{F}_o$, which *may differ* from the message format). libFTE also employs the RtE method, where regular expressions (regexes) represent formats, that are ranked using either a corresponding deterministic finite automaton (DFA) or non-deterministic finite automaton (NFA). More specifically, a DFA can be obtained through a general regex-to-DFA transformation, which is not always efficient. To allow the use of the (more efficient) regex-to-NFA transformation, the authors relax the ranking method, such that it can also be based on an NFA. Though presenting a general FPE scheme, the goals, focus, and solutions, of [9] are very different than ours. First, libFTE is designed for developers, and as such provides the developer with several possible schemes, out of which she chooses the most appropriate one. Our scheme is designed to be incorporated into a larger system which is designed for the end-user, so it must provide *a single* scheme, and the flexibility of our system is obtained by setting (according to the clients' specifications) few parameters "once and for all" in the larger, "wrapper" system. Second, formats in our scheme are defined directly, and naturally, from their user-defined properties, and is therefore flexible since the user can define new formats himself. Defining new formats in libFTE requires a developer's involvement to construct a regex from the user-defined format properties. This representation using regexes has the additional disadvantage of nonuniformity, since the performance of the resultant scheme depends on the *specific regex* chosen to represent the format, as opposed to the "complexity" of the format (as in our scheme). Moreover, there is no method of predicting whether the resultant scheme would have poor performance, and if it does, the developer cannot know whether a different regex would give better performance. We note that both our scheme, and libFTE, have the same security guarantees (since the underlying non-FPE scheme is the same in both). See Section V for performance comparison.

## II. PRELIMINARIES

NOTATION. The length of a string $s$ is denoted $|s|$, and $|A|$ denotes the number of elements in a set $A$. $x \cdot y$ denote the concatenation of strings $x, y$. We say that a string $s = c_1...c_n$ is *over alphabet* $\Sigma$ (where $\Sigma$ is a set of characters) if $\forall 1 \leq i \leq n, c_i \in \Sigma$. A format $\mathcal{F}$ defines a set of constraints over strings, i.e., it is associated with a set $\{s_1,...,s_n\}$ of strings. If $s = s_i$ for some $1 \leq i \leq n$ then we say $s$ *has format* $\mathcal{F}$ (denoted $s \in \mathcal{F}$). A format $\mathcal{F} = \{s_1,...,s_n\}$ is over alphabet $\Sigma$ if $\Sigma$ contains all the characters that appear in $s_1,...,s_n$. We say that $\mathcal{F}, \mathcal{F}'$, defined over alphabets $\Sigma, \Sigma'$ respectively, are *over disjoint alphabets* if $\Sigma \cap \Sigma' = \emptyset$. Similarly, a character $c$ is disjoint to format $\mathcal{F}$ if $c \notin \Sigma$, where $\mathcal{F}$ is defined over alphabet $\Sigma$.

### A. (Format-Preserving) Encryption Schemes.

An encryption scheme is a quadruple $\Pi = (\mathcal{M}, \mathsf{KeyGen}, \mathsf{Enc}, \mathsf{Dec})$, where $\mathcal{M}$ is a set of *plaintexts* (or *messages*), $\mathsf{KeyGen} : \mathbb{N} \to \mathcal{K}$ is a probabilistic polynomial-time algorithm that outputs an *encryption key* $k$ from the set $\mathcal{K}$ of keys, and $\mathsf{Enc} : \mathcal{K} \times \mathcal{M} \to \mathcal{C}, \mathsf{Dec} : \mathcal{K} \times \mathcal{C} \to \mathcal{M}$ are polynomial-time algorithms. Enc, on input an encryption key $k$ and a plaintext $m$, returns a ciphertext $c$; Dec, on input a key $k$ and ciphertext $c$, returns (the only) message $m$ such that $\mathsf{Enc}(k, m) = c$.

A *Format-Preserving* Encryption (FPE) scheme for format $\mathcal{M}$ is an encryption scheme with the additional property that $\mathcal{M} = \mathcal{C}$. Most FPEs studied in the literature are designed to encrypt only *specific* formats (e.g., credit-card numbers), while we focus on *General-Format* Preserving Encryption (GFPE), which can encrypt messages from various formats. More specifically, the encryption and decryption algorithms of a GFPE are associated with a collection of plaintext domains $\mathcal{M}_1,...,\mathcal{M}_n$. Given an encryption key $k$, a message $m \in \mathcal{M}_i$ (resp., ciphertext $c \in \mathcal{M}_i$), and a succinct representation $\mathsf{rep}(\mathcal{M}_i)$ of $\mathcal{M}_i$, encryption (resp., decryption) outputs a ciphertext $c \in \mathcal{M}_i$ (resp., the message $m \in \mathcal{M}_i$ such that $\mathsf{Enc}(k, m, \mathsf{rep}(\mathcal{M}_i)) = c$).

### B. Security Notions For FPEs.

Intuitively, encryption schemes should be "as unpredictable as possible", i.e., given a ciphertext an adversary should be unable to deduce *any properties of the encrypted message*, and this should hold *even* given prior knowledge on the message, and (possibly also) other ciphertexts encrypted using the same key. However, FPEs cannot achieve these security notions since they inherently reveal the message format. Consequently, the following four FPE-specific game-based security notions have been suggested [2]. *Pseudo-Random Permutation (PRP) security* requires that an adversary cannot distinguish encryptions with a randomly chosen key from random permutations over the format domain; *single-point indistinguishability (SPI)* requires that the adversary cannot distinguish the encryption of any message of its choice from a random ciphertext; *message privacy (MP)* requires that ciphertexts reveal no information on the encrypted message, except its format (this is formalized by comparing the "performance" of the real-world adversary to that of a degenerate adversary $\mathcal{S}$ that can only make equality queries of the form "is $m$ the encrypted message?"); and similar to MP, but weaker than it, *message recovery (MR)* only requires that the ciphertext does not *completely* reveal the encrypted message. The two latter security notions should hold even if the adversary can choose the message distribution to its advantage. (These security notions are non trivial since the degenerate adversary $\mathcal{S}$ operates on the same message distribution. For example, if the distribution is concentrated on one message, $\mathcal{A}$ has no advantage over $\mathcal{S}$ since both can recover the original message.) In all cases, the adversary is also given an encryption oracle. Roughly speaking, the advantage $\text{Adv}^{\mathsf{X}}(\mathcal{A})$ of an adversary $\mathcal{A}$ (where $\mathsf{X} \in \{\text{PRP}, \text{SPI}, \text{MP}, \text{MR}\}$) is the difference between the probability that $\mathcal{A}$ correctly guesses which situation he is in, and the probability of guessing correctly when only the format is known (in the first two cases, this probability is $\frac{1}{2}$). (Due to space limitations, we refer the interested reader to [2] for the formal definitions.) Bellare et al. [2] show that $\text{PRP} \to \text{SPI} \to \text{MP} \to \text{MR}$, meaning PRP is the strongest security notion and MR is the weakest. We note that though PRP is the best security notion one can hope to achieve for FPEs, the three weaker notions can, in many *concrete* cases, offer better security for the same efficiency, and may therefore suffice in practice.

## III. Existing GFPE Schemes and Their Limitations

Cycle-walking and RtE provide general strategies for constructing GFPEs, but a GFPE also requires *efficient* ranking and unranking methods. Such methods were described for a specific format type [13], [1], and were claimed to apply to more general formats. In this section we review and analyze these methods, and show that applying them to more general formats yield completely insecure schemes.

### A. SGFPE: A Simple General-Format FPE

In this section we review the GFPE scheme of [13], [1], which we call SGFPE. At a high-level, SGFPE first "simplifies" a given format $\mathcal{F}$ by replacing it with another format $\mathcal{SF}$(described below), which usually consists of partitioning it into smaller sub-formats ("shrinking"), but may also add strings to $\mathcal{F}$ ("expansion"). Strings are ranked in $\mathcal{SF}$ by applying a method similar to counting (together with cycle-walking, if $\mathcal{SF}$ strictly contains $\mathcal{F}$). SGFPE can encrypt any format whose simplified form consists of fixed-length strings with independent location-specific character-sets. That is, $\mathcal{SF}$ is defined by a length $\ell$, and $\ell$ sets of characters $\Sigma_1, ..., \Sigma_\ell$, and contains all $\ell$-length strings $s = c_1...c_\ell$ such that $c_i \in \Sigma_i$ for every $1 \leq i \leq \ell$. A format $\mathcal{F}$ can be simplified by partitioning it to the union of simplified formats $\mathcal{SF}_1, ..., \mathcal{SF}_m$ (in which case a string $s$ is ranked using the ranking method of $\mathcal{SF}_i$, where $s \in \mathcal{SF}_i$). Alternatively, $\mathcal{SF}$ can be obtained by removing restrictions defining $\mathcal{F}$. (For example, the format $\mathcal{F}$ of all legal social-security numbers can be simplified to the format $\mathcal{SF}$ of all 9-digit decimal strings, removing any additional restrictions $\mathcal{F}$ imposes on legal social-security numbers.) A simplified format $\mathcal{SF}$ is ranked using the *Scale-and-Sum* (SaS) method (see Section IV-C1).

### B. Limitations of SGFPE

*1) (In)security Of SGFPE - Theoretical Analysis:* We analyze SGFPE security according to the FPE security notions [2]. Since MR is the weakest (see Section II-B), showing that SGFPE is insecure according to MR implies its insecurity according to *all* four security notions. Remember that SGFPE partitions a format $\mathcal{F}$ into sub-formats $\mathcal{SF}_1, ..., \mathcal{SF}_k$, and encrypts a message $m$ to a ciphertext $c$ in the same sub-format. Therefore, $c$ reveals the sub-format $\mathcal{SF}_i$ to which $m$ belongs, thus decreasing the number of possible messages, and increasing the probability of recovering $m$.[1] We formalize this intuition by describing an adversary that breaks SGFPE MR-security. We first describe an adversary for sparse formats (i.e., formats in which message-specific characteristics almost define the message), and then extend the attack to dense formats. Bellare et al. [2] show that for an adversary $\mathcal{A}$ making $q$ oracle queries and choosing message distribution $\mathcal{D}$, the best strategy of the degenerate adversary $\mathcal{S}$ is to make equality queries on $m_1, ..., m_q$, where $m_i \neq m_j$ for every $i \neq j$, and for every $1 \leq i \leq q$, if $m_1, ..., m_{i-1}$ are discarded then $m_i$ is the most likely message (according to $\mathcal{D}$). If $m \in \{m_1, ..., m_q\}$, then $\mathcal{S}$ outputs $m$, otherwise he guesses that $m_{q+1}$ (i.e., the likeliest message once $m_1, ..., m_q$ are discarded) is the encrypted message. The probability that $\mathcal{S}$ recovers $m$ is $\sum_{i=1}^{q+1} \Pr_\mathcal{D}[m = m_i]$.

ATTACKING SPARSE FORMATS. Consider the format $\mathcal{F} = \{m_1, m_2, ..., m_k\}$, where $\forall 1 \leq i \leq k, |m_i| = \ell_i$, and $\ell_i \neq \ell_j$ for every $i \neq j$. Let $\mathcal{A}$ be an adversary that chooses the uniform distribution on $\mathcal{F}$, and makes no oracle queries. Then $\text{Adv}^{\text{MR}}(\mathcal{A}) = 1 - \frac{1}{k}$. Indeed, the ciphertext length reveals the message so $\mathcal{A}$ always guesses correctly. The optimal $\mathcal{S}$ picks a random message in $\mathcal{F}$ (because he cannot make any oracle queries), recovering $m$ with probability $\frac{1}{k}$.

ATTACKING GENERAL FORMATS. The previous attack succeeded because in sparse formats the ciphertext length reveals the message. However, an adversary $\mathcal{A}_{\text{MR}}$ can "transform" *any* (non-sparse) format $\mathcal{F}$ into a sparse one, as follows. $\mathcal{A}_{\text{MR}}$ partitions $\mathcal{F}$ into sub-formats $\mathcal{SF}_1, ..., \mathcal{SF}_k$, where for every $1 \leq i \leq k$, all messages in $\mathcal{SF}_i$ have the same length $\ell_i$, and $\ell_i \neq \ell_j$ for $i \neq j$. Then, $\mathcal{A}_{\text{MR}}$ chooses $k$ arbitrary messages $m_1 \in \mathcal{SF}_1, ..., m_k \in \mathcal{SF}_k$, and takes $\mathcal{D}$ to be the uniform distribution over $\{m_1, ..., m_k\}$. This effectively reduces $\mathcal{F}$ to a sparse format, so the previous analysis shows that $\text{Adv}^{\text{MR}}(\mathcal{A}_{\text{MR}}) = 1 - \frac{1}{k} \to^{k \to \infty} 1$, i.e., the advantage increases with the format diversity (i.e., with the variety of message lengths).

---

[1]In general, smaller format are more predictable and therefore more prone to attacks (e.g., an attacker correctly guesses a value in a gender column with probability $\frac{1}{2}$, and statistics about the male-female ratio in the organization can drastically improve this probability).

In summary, we have shown an adversary breaking the MR-security of SGFPE, and thus also the MP-, SPI- and PRP-security of SGFPE. The advantage can be made arbitrarily close to 1 as the format in question is more diverse.

*Example 3.1:* Consider the format $\mathcal{F}_{\text{name}}$ of names, consisting of 1-4 words, where each word has the format $\mathcal{F}_{\text{word}}$ of an upper-case letter, followed by 0-63 lower-case letters, followed by a space. For $x \in \mathcal{F}_{\text{name}}$ (for $w \in \mathcal{F}_{\text{word}}$), let $N_{\text{word}}(x)$ ($N_{\text{letter}}(w)$) denote the number of words (letters) in $x$ ($w$). Here, each sub-format is defined by $N_{\text{word}}, N_{\text{letter}}$, so $k = \sum_{i=1}^{4} 64^i$, and $\text{Adv}^{\text{MR}}(\mathcal{A}) = 1 - \frac{1}{\sum_{i=1}^{4} 64^i}$.

*2) (In)security Of SGFPE - Practical Analysis:* We draw on the theoretical weaknesses discussed in Section III-B1, and design practical attacks that breach the data privacy, thus rendering SGFPE impractical from a security standpoint.

ATTACK MODEL. Our attack model is motivated by the common scenario of third-party hosting with external attacks, e.g., when encrypted databases are stored at remote servers. A database may be accessed both by legitimate applications, and by unauthorized parties. Thus, an unauthorized party (adversary) may have partial or complete knowledge of the database. However, as the remote server does not know the secret encryption key being used, it is unlikely that the adversary can learn the encryption of messages of its choosing. To capture these adversarial capabilities, we assume the adversary is given full access to the encrypted database. In addition, we assume the adversary has some prior knowledge $\mathcal{I}$ regarding the data (e.g., the adversary may know the database contains medical records of residents of a specific city). The goal of the adversary is to extract as much information as possible, about as many records as possible. This adversarial goal was chosen to reflect real-world threats, as researches show that even few properties of an individual completely identify her (see, e.g., [12]). Therefore, we consider privacy to be broken, even if the adversary is able to "only" pin-point a few specific properties of a certain record.

RESULTS. Our experiments were performed on the FEC reports. (The Federal Election Commission (FEC) is a regulatory agency that regulates the campaign finance legislation in the United States.) These reports list all donors (that donated more than $200), along with the name, home address, employer and job title of each donor. Experiments were performed on the first 1,000,000 records out of the 2008-2012 report, available at https://explore.data.gov/Contributors/FEC-Contributions/4dkz-64bn. Our results show that dataset records may be clustered into groups, based on message-specific attributes preserved by SGFPE, e.g., the number of words in every column, and the number of letters in every word. (The graphs show, for every percentage of the population, a lower bound on the probability with which it can be identified. For example, if 50% of the records can be identified with probability 0.5, then for 50% of the records, their encryption is consistent with only 2 unencrypted records.) Concretely, we show (Figure 1, left) that if the FEC report is encrypted using SGFPE then 71% of the donors can be identified with probability at least 0.5, and another 15% can be identified with probability at least 0.1. (This means that the identifiers of 71% of the donors match only two records.) We also analyzed the level of security guaranteed if the adversary could access only part of the table. For example, we show (Figure 1, right) that if only the name and town columns are revealed then 16% of the donors can be identified with probability at least 0.1, and another 28% can be identified with probability at least 0.01. (Additional graphs can be found in the full version.) This security loss should be contrasted to the level of security achievable by a non-FPE scheme, which resembles a random permutation (see Section II-B).

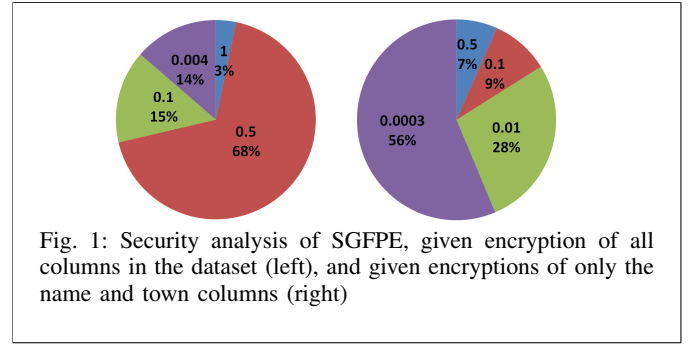

Fig. 1: Security analysis of SGFPE, given encryption of all columns in the dataset (left), and given encryptions of only the name and town columns (right)

*3) SGFPE - Efficiency Analysis:* We analyze the concrete efficiency of SGFPE. As the "costly" operations are encryption and decryption, we measure efficiency in terms of the number of integer-FPE encryption and decryption operations when encrypting a single message (i.e., the cycle length during encryption or decryption). Our approach is further motivated by the following observation. When basing general-format FPE on the RtE framework, each of the three operations (ranking, encrypting or decrypting, and unranking) is performed at least once, so repeating any of these operations is considered redundant, but ranking and unranking are never repeated. As noted in Section III-A, when a format $\mathcal{F}$ is simplified by expansion to format $\mathcal{SF}$ then cycle-walking is used to guarantee that the ciphertext is in $\mathcal{F}$. The average cycle length is $\frac{|\mathcal{SF}|}{|\mathcal{F}|}$, so reducing the ratio improves average-case efficiency, but also degrades security (and is sometimes impossible, as in the case of social-security numbers). Consequently, the average cycle-length may be long. (See [9] for additional examples.)

*Example 3.2:* consider the format $\mathcal{F}$ of strings of the form date, ssn, ccn, where date is a date between 01.01.1900 and 23.09.2013, ssn is a valid social security number, and ccn is a 16-digit credit card number, where the $16^{\text{th}}$ digit is a luhn check-sum digit. ($\mathcal{F}$ can be used to describe credit-card transactions, by providing the card holder's SSN and the date of the transaction.) $\mathcal{F}$ can be represented in SGFPE as all strings of the form $x_1x.y_1y.z_1zzz, w^9, c^{16}$ where $x_1 \in \{0,1,2,3\}$, $y_1 \in \{0,1\}, z_1 \in \{1,2\}, x,y,z,w,c \in \{0,1,...,9\}$ (where $n^m$ denotes a sequence of $m$ characters of type $n$, not necessarily identical). In this case, the average cycle-length is more than 629 (i.e., integer-FPE encryption is repeated more than 629 times on average *when encrypting a single message*). We note that the average cycle-length can be shortened by splitting the format into sub-formats, but this will decrease the security of the scheme. For example, we can divide $\mathcal{F}$ into sub-formats $\mathcal{F}_1, \mathcal{F}_2$, where $x_1x < 30$ for all messages $m \in \mathcal{F}_1$, and $x_1x \in \{30, 31\}$ for all messages $m \in \mathcal{F}_2$. This reduces the average cycle-length by a factor of 4, but significantly degrades security since a ciphertext with prefix $3x.yy.zzzz$ could only result from a transaction that took place on the $30^{\text{th}}$ or the $31^{\text{st}}$. The alternative solution of encrypting message ingredient (date, SSN, CCN) separately also substantially degrades security, due to the small size of the sub-formats

More generally, the average encryption time is $t_{\text{Enc}} = t_{\text{rank}} + AL_{\text{cy}} \cdot t_{\text{intEnc}} + t_{\text{unrank}}$, where $t_{\text{rank}}, t_{\text{intEnc}}, t_{\text{unrank}}$ denote the running time of the ranking, integer-FPE encryption and unranking algorithms, and $AL_{\text{cyc}}$ is the average cycle length. However, $\frac{|S\mathcal{F}|}{|\mathcal{F}|}$ is an *average* bound (the *actual* length may be larger, and cannot be predicted), so in many cases it is highly desirable to completely eliminate cycle-walking.

## IV. OUR GFPE SCHEME

### A. Overview

The main shortcoming of SGFPE (Section III) is its inflexibility in format representation: it offers a single, very specific method of representing general formats, focusing on a specific set of properties (length and location-specific character-sets), while ignoring all other format properties. As we have shown, this results in a scheme which is insecure and achieves non-optimal efficiency. Our scheme is also based the RtE framework, but by providing a flexible framework of representing general formats, we improve security and efficiency.

REPRESENTING FORMATS. We cannot possibly predict all formats to which our GFPE scheme may be applied, so we supply several format "building-blocks", from which compound formats are constructed by applying "composition operations" which we define. Thus, we get a well-defined method of representing compound formats admitting simpler ranking and unranking operations. More specifically, the building blocks, or *Primitives*, represent non-compound formats (usually "rigid", i.e., defined by a strict, non-flexible set of rules). Primitives sometimes require specially-tailored ranking and unranking algorithms. Using the composition operations, compound formats, or *Fields*, can be constructed from any set of primitives or fields. Intuitively, the composition operations preserve the property that a message string $s$ in a compound field-format $\mathcal{FL}$, which was constructed from primitives and fields $\mathcal{FL}_1, ..., \mathcal{FL}_k$, can be *efficiently* parsed into substrings $s = s_1...s_k$, such that $\forall 1 \leq i \leq k, s_i \in \mathcal{FL}_i$. (We refer to this property as the *parsability* property.) Thus, after parsing the string, ranking and unranking can be (mostly) delegated to the ingredients of the compound field.

RANKING AND UNRANKING. We provide *efficient* ranking and unranking methods for all formats representable in our framework. Concretely, we provide ranking and unranking algorithms for all primitives and composition operations. Thus, primitives are ranked directly; and compound fields can be ranked using the ranking method of the composition operations with which they were constructed, and by delegating (some of) the work to the fields and primitives from which they were constructed.

ENCRYPTION AND DECRYPTION. As our system uses RtE, encryption and decryption of formats reduce to integer-FPE encryption and decryption. Our scheme has the flexibility that it can use either FE1, FE2 [2], or FFX [3]. FE1 and FE2 completely eliminate cycle walking, and their security has been rigorously analyzed in [2]. However, implementations of these algorithms are currently less efficient than FFX, because they employ factorization algorithms. FFX must be used together with cycle-walking since, strictly speaking, it is not an integer-FPE. Recall that we cannot bound the *actual* cycle-length, but for an appropriate choice of the parameters of FFX we can guarantee that the *average* cycle length is at most 2. Currently, FFX implementations are faster than FE1,FE2. FFX has no rigourous security analysis, but is currently under consideration of NIST, and may become an FPE standard.

### B. Representing General Formats

We tried to identify the minimal set of primitive formats and composition operations from which general formats can be constructed, and focused on formats and operations that admit simple and efficient ranking and unranking algorithms. We stress that the main advantage of our scheme, compared to SGFPE, is that it allows one to represent complex formats by identifying their properties and ranking a message in relation to *all, and only,* messages with the same format.

*Remark 4.1:* There may exist some specific non-compound formats we have not incorporated into our framework. If needed, these can be added by presenting ranking and unranking methods for these formats. Thus, our framework is also easily extendable and adjustable.

*1) Primitive (Non-Compound) Formats:* We present several primitive formats, which serve as building-blocks for constructing compound formats. We distinguish between *rigid* primitives $\mathcal{RP}$, possessing the *prefix-parsability* (which is used to preserve the parsability property when defining operations), namely given a string $s = s's''$ such that $s'$ has format $\mathcal{RP}$, $s'$ can be efficiently parsed from $s$ in one pass; and *non-rigid* formats, that do not posses this property. Ranking and unranking primitives usually require specially-tailored algorithms (due to space limitation, we defer most of these algorithms to the full version). In the following, all primitives are rigid, unless specifically noted otherwise.

SOCIAL SECURITY NUMBERS (SSNS). A 9-digit decimal number $x_9x_8x_7x_6x_5x_4x_3x_2x_1$, where $x_9x_8x_7 \notin \{000, 666\} \land x_9x_8x_7 < 900$, $x_6x_5 \neq 00$ and $x_4x_3x_2x_1 \neq 0000$.

CREDIT CARD NUMBERS (CCNS). A 16-digit decimal string, where the $16^{\text{th}}$ digit is a luhn sum-check digit of the first 15 digits.

DATES. A set of legal dates is defined by minimal and maximal dates $\text{minD}, \text{maxD}$, whose format determines the granularity (e.g., whether dates are of the form "$dd.mm.yyyy$" or "$dd.mm.yyyy\ hh : mm : ss$"). The minimal date is needed for ranking, since ranks are computed in relation to it (see Section IV-C). The maximal date is needed since the format must be finite. (We cannot use the current date as a maximal date, since future dates may also be encrypted.)

FIXED-LENGTH STRINGS. $\ell$-length strings (for some fixed length $\ell$), where for every $1 \leq i \leq \ell$, the character in location $i$ is from a specific set $\Sigma_i$ of characters. (This is the format captured by SGFPE.)

DELIMITERED VARIABLE-LENGTH STRINGS. Defined by minimal and maximal lengths $\text{min}, \text{max}$, a set (alphabet) $\Sigma$ of legal characters, and a delimiter character $d \notin \Sigma$. It consists of all strings of the form $s \cdot d$ where $s$ is over alphabet $\Sigma$, and $\min \leq |s| \leq \max$. (This format can be generalized to delimited strings over location-specific character sets. However, as this generalized formats seems not useful in practice, we consider only the simpler form.)

NON-DELIMITED VARIABLE-LENGTH STRINGS. The non-rigid equivalent of the "delimited variable-length strings" format, it is defined by the same parameters $\min, \max, \Sigma$. It consists of all strings $s$ over alphabet $\Sigma$ such that $\min \leq |s| \leq \max$. (This format can also be generalized to location-specific character sets.)

DELIMITED LEGAL STRINGS-SET. The set of strings defines the only "legal" strings, i.e., the only strings having the format defined by the set. To achieve the prefix-parsability guarantee, we require that all strings in the set be delimited by a character $d$ that does not appear in any other location in the string. We also extend this primitive to *prefix-free* sets of (not necessarily) delimited strings. (By prefix-free we mean that there exist no $s, s'$ in the set such that $s$ is a prefix of $s'$.) Though every format can be represented by a set of legal strings, ranking and unranking here must keep translation tables, and are therefore practical only for very small sets, so this primitive should only be used when the format cannot be defined by any other property (see Example 4.2 below).

LEGAL STRINGS-SET. The non-rigid equivalent of the delimited legal strings-set format, in which the set of legal strings need not be prefix-free or delimited.

INTEGRAL DOMAIN. The set of integers between some min and max values. Though this format is trivial (since it is a domain of integers and therefore can be encrypted directly using an integer-FPE), we incorporate it into our framework so that integral domains can be used as building-blocks for designing compound formats (see Example 4.2 below).

*2) Composition Operations:* Our framework includes three composition operations, that maintain the parsability property discussed in Section IV-A. Consequently, ranking and unranking can be performed by delegating the ranking to underlying sub-formats, as we show in Section IV-C below.

UNION. The format $\mathcal{F}$ is defined as the union of *disjoint* formats $\mathcal{F}_1, ..., \mathcal{F}_k$, and consists of all strings having one of the formats $\mathcal{F}_1, ..., \mathcal{F}_k$.

CONCATENATION. The format $\mathcal{F}$ can be defined by the concatenation of formats $\mathcal{F}_1, ..., \mathcal{F}_k$. The concatenation can be performed in one of two ways, depending on the properties of the underlying formats $\mathcal{F}_1, ..., \mathcal{F}_k$. (The difference is due to the parsability property which the concatenation should preserve.)

- $\mathcal{F} = \mathcal{F}_1 \cdot ... \cdot \mathcal{F}_k$ (i.e., every string $s \in \mathcal{F}$ can be written as $s = s_1...s_k$ where $\forall 1 \leq i \leq k, s_i \in \mathcal{F}_i$), such that for every $1 \leq i \leq k-1$, $\mathcal{F}_{i+1}$ is separable from $\mathcal{F}_i$, where separability is defined as follows. $\mathcal{F}$ is separable from $\mathcal{F}'$ either if $\mathcal{F}'$ is a *rigid* primitive, or if $\mathcal{F}, \mathcal{F}'$ are defined over disjoint alphabets.

- $\mathcal{F} = \mathcal{F}_1 \cdot d_1 \cdot \mathcal{F}_2 \cdot d_2 ... \cdot d_{k-1} \cdot \mathcal{F}_k$ (i.e., every string $s \in \mathcal{F}$ can be written as $s = s_1 d_1 s_2 d_2 ... d_{k-1} s_k$ where $\forall 1 \leq i \leq k, s_i \in \mathcal{F}_i$), where $d_1, ..., d_{k-1}$ are characters (called *delimiters*, since they are used to separate $s_i$ from $s_{i+1}$), and for every $1 \leq i \leq k-1$, $d_i$ is disjoint to $\mathcal{F}_i$. Here, the parsability property is maintained since upon reaching $d_i$, the parser knows that the previous character is the last character of $s_i$.

RANGE. Analogously to the "variable-length strings" primitive, a format $\mathcal{F}$ can be defined as a variable-length concatenation of a sub-format $\mathcal{F}'$. Concretely, $\mathcal{F}$ is defined by $\mathcal{F}'$, a delimiter character $d$ that is disjoint to $\mathcal{F}'$, and a pair of numbers $\min < \max$. $\mathcal{F}$ contains all strings $s = s_1 \cdot d \cdot s_2 \cdot d ... d \cdot s_k \cdot d$ such that $\forall 1 \leq i \leq k, s_i \in \mathcal{F}'$, and $\min \leq k \leq \max$. (We also consider a variant of the range operation, in which the final substring $s_k$ is not delimited.)

*Example 4.2 (addresses):* Consider the format $\mathcal{F}_{\text{add}}$ of legal addresses of the form "$\text{name}_1$ num $\text{name}_2$ $\text{name}_3$ zip state", where $\text{name}_1, \text{name}_2, \text{name}_3 \in \mathcal{F}_{\text{name}}$ are the name, street name and city name, respectively (1-4 words, each an uppercase letter followed by 0-63 lowercase letters and a space); num is the street number (say, a number between 1 and 1053); zip is a zip-code (a 5-digit decimal number); and state is a 2-character state code. Then $\mathcal{F}_{\text{add}}$ can be represented as the concatenation $\mathcal{F}_{\text{add}} = \mathcal{F}_{\text{name}} \cdot \mathcal{F}_{\text{num}} \cdot \mathcal{F}_{\text{space}} \cdot \mathcal{F}_{\text{dName}} \cdot \mathcal{F}_{\text{zip}} \cdot \mathcal{F}_{\text{space}} \cdot \mathcal{F}_{\text{state}}$, where $\mathcal{F}_{\text{name}}$ ($\mathcal{F}_{\text{dName}}$) is the range composition with parameters $\mathcal{F}_{\text{word}}, ' ', 1, 4$ ($\mathcal{F}_{\text{word}}, ' ', 2, 8$), and $\mathcal{F}_{\text{word}}$ is a "variable-length strings" primitive; $\mathcal{F}_{\text{num}}$ is the (shifted) integral domain $\{1, 2, ..., 1053\}$; $\mathcal{F}_{\text{space}}, \mathcal{F}_{\text{zip}}$ are "fixed-length string" primitives; and $\mathcal{F}_{\text{state}}$ is a "delimited legal string-set" primitive.

*Example 4.3 (credit-card transactions):* The format of Example 3.2 can be represented as the concatenation $\mathcal{F}_{\text{trans}} = \mathcal{F}_{\text{date}} \cdot \mathcal{F}_{\text{com}} \cdot \mathcal{F}_{\text{ssn}} \cdot \mathcal{F}_{\text{com}} \cdot \mathcal{F}_{\text{ccn}}$ where each of the sub-formats is a rigid primitive ($\mathcal{F}_{\text{com}} = \{","\}$). Unlike the representation in SGFPE (Example 3.2), this representation eliminates cycle-walking. (Further examples are deferred to the full version.)

### C. Ranking And Unranking General Formats

To describe how to *efficiently* rank and unrank general formats, we first describe the Scale-and-Sum method.

*1) The (Generalized) Scale-and-Sum (SaS) Method:* This is a generalization of the decimal counting method for ranking of strings. Let $\mathcal{F}$ be defined by a length $\ell$ and location-specific character sets $\Sigma_1, ..., \Sigma_\ell$, and let $N_i := |\Sigma_i|, i = 1, ..., \ell$, and $O_i : \Sigma_i \to N_i, i = 1, ..., \ell$ be arbitrary orderings of $\Sigma_1, ..., \Sigma_\ell$. (Notice that storing these orderings is significantly cheaper than storing an ordering of the entire format, since the $N_i$'s are small.) Then $\text{rank}(s) = \sum_{i=1}^{\ell} O_i(c_i) \cdot \prod_{j=1}^{i-1} N_j$ for $s = c_1...c_\ell \in \mathcal{F}$ is the number of strings before $s$ in the lexicographic order on $\mathcal{F}$, the set of length-$\ell$ strings where every $i^{\text{th}}$ character is in $\Sigma_i$. In this sum, characters have "weights" according to their location (just as binary strings have *most* and *least* significant bits), and so the location $O_i(c_i)$ of the character $c_i$ in the set $\Sigma_i$ is scaled by its weight (i.e., the number of length-$(i-1)$ strings in which every character $c_j$ is from the set $\Sigma_j$). The operation $\text{unrank} := \text{rank}^{-1}$ retrieves the characters of $s$ one-by-one (just as taking a decimal number modulo 10 retrieves its digits). More specifically, the location $O_i(c_i)$ of a character $c_i$ in the set $\Sigma_i$ is retrieved by first removing the "contributions" of the characters $c_1, ..., c_{i-1}$, and taking the result modulo $\prod_{j=1}^{i-1} N_i$.

(By retrieving the characters $c_1, ..., c_\ell$ from left-to-right, the "contribution" $\sum_{j=1}^{i-1} O_j(c_j) \cdot \prod_{l=1}^{j-1} N_l$ of $c_1, ..., c_{i-1}$ can be computed and removed from $\text{rank}(s)$.) This method extends to variable-length strings by incorporating the number of shorter strings into the rank. For example, in the format of all strings $s = c_1...c_\ell$ of length $\ell_1 \leq \ell \leq \ell_2$ such that $c_i \in \Sigma_i$ for every $1 \leq i \leq \ell$, $\text{rank}(s) = \sum_{i=\ell_1}^{i=\ell-1} \prod_{j=1}^{j=i} N_j + \sum_{i=1}^{\ell} O_i(c_i) \cdot \prod_{j=1}^{i-1} N_j$, where $|s| = \ell$ (the first summand is the number of strings in $\mathcal{F}$ whose length is shorter than $\ell$). As we show in Section IV-C3, the SaS method can be generalized to complex sets $\Sigma_i$ containing *strings* instead of characters, and will be a main tool in ranking and unranking operations.

*2) Ranking And Unranking Primitive Formats:* SOCIAL-SECURITY NUMBERS (SSNs). Disregarding the specific restrictions on valid SSNs, they are simply 9-digit decimal strings, and so the rank of a given SSN $s$ is itself. As valid SSNs *do* follow certain rules, the ranking method subtracts from $s$ the number $|\{n : n < s \wedge n \text{ is not a valid SSN}\}|$. This is done by distinguishing between SSNs that are smaller than $666xxxxxx$, and SSNs that are larger than $666xxxxxx$. (This distinction is required because all numbers of the form $666xxxxxx$ are not valid SSNs, and should be subtracted *only* from the rank of the *second* type.) The unranking procedure, given a rank $r$, should retrieve the SSN $s$ such that $\text{rank}(s) = r$, namely it should return $r + N_s$, where $N_s := |\{n : n < s \wedge n \text{ is not a valid SSN}\}|$. Since $s$ is unknown, a straight-forward computation of $N_s$ is rather complex. Therefore, we implement the unranking method by using a variant of binary search on the message domain of legal SSNs. Further details are deferred to the full version.

CREDIT CARD NUMBERS (CCNs). The rank of $c = c_1 c_2 ... c_{16}$ is the 15-digit decimal number $c_1 c_2 ... c_{15}$. To unrank $r = \text{rank}(c)$ we interpret $r$ as a 15-digit string $r_1 r_2 ... r_{15}$ (adding leading zeros if needed), and concatenate to it from the right the luhn sum-check digit of $r_1, r_2, ..., r_{15}$.

DATES. We discuss ranking for dates of the form "$dd.mm.yyyy\ \ hh : mm : ss$" (dates of the form "$dd.mm.yyyy$" are ranked similarly by counting days instead of seconds). The rank $r$ of a date $d$ is the number of seconds $N$ since minD. $N$ can be computed by first determining the relative number of seconds, minutes, hours etc. of $d$ (compared to minD), and then computing $N$ according to the SaS method (where every element is scaled by the number of seconds in a "unit" of that element, e.g., the number of days is scaled by the number of seconds in a day). To unrank $r = \text{rank}(d)$, we first retrieve the relative number of seconds, minutes and hours (compared to minD), which can be computed by modular operations. Similarly, we can retrieve the *total* number of days passed since minD, and taking into consideration leap years etc., we can compute the relative number of years, months and days. The non-relative date $d$ can then be retrieved from minD and the relative values.

FIXED-LENGTH STRINGS. Uses the SaS method.

DELIMITERED AND NON-DELIMITERED VARIABLE-LENGTH STRINGS. Ranking uses the *extended* SaS method (in the delimitered case, the delimiter character is ignored during ranking). Given a rank $r = \text{rank}(s)$, unranking first determines $\ell = |s|$ (this can be done since $N_\ell \leq r < N_{\ell+1}$ where $N_i$ is the number of strings of length $\leq i$ in the format). Then, $s$ is retrieved from $r' = r - N_\ell$ (which is the rank of $s$ among length-$\ell$ strings in the format) using the fixed-length strings unranking algorithm. (In the delimitered case, $s$ is retrieved *without* the delimiter, which is then concatenated to it.)

DELIMITERED AND NON-DELIMITERED LEGAL STRINGS-SET. Through search in look-up tables.

INTEGRAL DOMAIN. Ranking and unranking shift the given number n by the minimial range point.

*3) Ranking And Unranking Compound Formats:* At a high level, a string $s$ in a compound format $\mathcal{F}$ is ranked by parsing it into substrings, where each substring is in a sub-format used to define $\mathcal{F}$ (parsing is done according to the composition operation used to construct $\mathcal{F}$), and delegating the ranking task to these sub-formats. Then, $\text{rank}(s)$ is constructed from the ranks in the sub-format using the *generalized* SaS method, where the scaler of every sub-format $\mathcal{F}_i$ is $\prod_{j=1}^{i-1} |\mathcal{F}_j|$. (This is a generalization in the sense that the location-specific sets $\Sigma_i$ are now sets of strings.) We now describe specifically how generalized SaS applies to compound formats, according to the composition operations used to define them.

UNION. Let $\mathcal{F} = \mathcal{F}_1 \cup ... \cup \mathcal{F}_k$. The ranking algorithm on input $s \in \mathcal{F}$ finds the index $1 \leq i \leq k$ such that $s \in \mathcal{F}_i$, and returns $\text{rank}_{\mathcal{F}_i}(s) + \sum_{j=1}^{i-1} |\mathcal{F}_j|$, where $\text{rank}_{\mathcal{F}_i}(s)$ is the rank of $s$ in the format $\mathcal{F}_i$. The unranking algorithm on input a rank $r = \text{rank}(s)$ operates as follows. First, it determines the smallest $1 \leq i \leq k$ such that $s \in \mathcal{F}_i$. (This can be done since $i$ is the smallest index such that $\sum_{j=1}^{i-1} |\mathcal{F}_j| \leq r < \sum_{j=1}^{i} |\mathcal{F}_j|$.) Once $i$ is known, $\text{rank}_{\mathcal{F}_i}(s)$ is computed and $s$ is found by applying the unranking operation of $\mathcal{F}_i$ to $\text{rank}_{\mathcal{F}_i}(s) = \text{rank}(s) - \sum_{j=1}^{i-1} |\mathcal{F}_j|$.

CONCATENATION. We describe ranking and unranking of formats $\mathcal{F} = d_1 \cdot \mathcal{F}_2 \cdot d_2 ... \cdot d_{k-1} \cdot \mathcal{F}_k$ obtained through concatenation, the algorithms for formats $\mathcal{F} = \mathcal{F}_1 \cdot ... \cdot \mathcal{F}_k$ are similar. On input $s \in \mathcal{F}$, the ranking algorithm parses it to $s = s_1 d_1 s_2 d_2 ... d_{n-1} s_n$ and computes $r_i = \text{rank}_{\mathcal{F}_i}(s_i)$, $i = 1, 2, ..., n$ (by delegating the ranking of $s_1, ..., s_n$ to $\mathcal{F}_1, ..., \mathcal{F}_n$). Then, the rank is computed using the generalized SaS method, i.e., $\text{rank}(s) = \sum_{i=1}^{n} r_i \cdot \prod_{j=1}^{i-1} |\mathcal{F}_j|$. The unranking algorithm on input $r = \text{rank}(s)$ uses the generalized SaS method to retrieve the ranks $r_1, ..., r_n$ of the substrings $s_1, ..., s_n$, then delegates the unranking of $r_1, ..., r_n$ to the sub-formats $\mathcal{F}_1, ..., \mathcal{F}_n$, and given their outputs $s_1, ..., s_n$, outputs the string $s = s_1 d_1 s_2 d_2 ... d_{n-1} s_n$.

RANGE. Let $\mathcal{F}$ be a range format defined with parameters $\mathcal{F}', d$ and $\text{min} < \text{max}$. Given a string $s \in \mathcal{F}$, ranking parses $s = s_1 d s_2 d ... s_k d$ for some $\text{min} \leq k \leq \text{max}$, and outputs $\text{rank}(s) = N_{k-1} + R_k$, where $N_{k-1} = \sum_{i=\text{min}}^{k-1} |\mathcal{F}'|^i$ is the number of strings consisting of $k'$ concatenations of substrings from $\mathcal{F}'$ (for some $\text{min} \leq k' \leq k-1$), and $R_k$ is the rank of $s$ among strings with *exactly* $k$ concatenations ($R_k$ is computed using the generalized SaS method as described above for a compound format obtained through concatenation). The unranking method, given a rank $r = \text{rank}(s)$ first determines $k$ (i.e., the index such that $N_{k-1} \leq r < N_k$), then computes $r' = r - N_{k-1}$ (namely, the rank of $s$ between strings with $k$ concatenations), and uses the generalized SaS method to retrieve from $r'$ the ranks $r_1, ..., r_k$ (of the substrings $s_1, ..., s_k$ defining the concatenation) in relation to the format $\mathcal{F}'$.

Finally, the substrings $s_1, ..., s_k$ are computed by delegating the unranking of $r_1, ..., r_k$ to $\mathcal{F}'$, and $s = s_1 d s_2 d ... s_k d$ is returned.

*Example 4.4 (ranking addresses):* Consider the format $\mathcal{F}_{\text{add}}$ (Example 4.2), To rank an address $s =$ "Jane Doe 53 Cherry Tree Road New York 12345 NY", it is first parsed as $s = s_1...s_7$ where $s_1 =$ "Jane Doe ", $s_2 =$ "53", $s_3 = s_6 =$ " ", $s_4 =$ "Cherry Tree Road New York ", $s_5 =$ "12345" and $s_7 =$ "NY". The ranks $r_1, ..., r_7$ of $s_1, ..., s_7$ (respectively) are computed using the ranking methods of $\mathcal{F}_{\text{name}}$, $\mathcal{F}_{\text{num}}$, $\mathcal{F}_{\text{space}}$, $\mathcal{F}_{\text{dName}}$, $\mathcal{F}_{\text{zip}}$ and $\mathcal{F}_{\text{state}}$. The rank of $s$ is then taken to be $r = r_1 + r_2 \cdot |\mathcal{F}_{\text{name}}| + r_3 \cdot |\mathcal{F}_{\text{name}}| \cdot |\mathcal{F}_{\text{num}}| + ... + + r_7 \cdot |\mathcal{F}_{\text{name}}| \cdot |\mathcal{F}_{\text{num}}| \cdot |\mathcal{F}_{\text{space}}|^2 \cdot |\mathcal{F}_{\text{doueblName}}| \cdot |\mathcal{F}_{\text{zip}}| \cdot |\mathcal{F}_{\text{state}}|$ To unrank $r = \text{rank}(s)$, the ranks $r_1, ..., r_7$ are retrieved from $r$ (e.g., $r_1 = r \mod |\mathcal{F}_{\text{name}}|$, and $r_2 = ((r \mod |\mathcal{F}_{\text{name}}|) - r_1) \mod |\mathcal{F}_{\text{num}}|$), and $r_1, ..., r_7$ are unranked to $s_1, ..., s_7$ using the unranking methods of the sub-formats. The output is the concatenation $s = s_1...s_7$.

### D. Security And Efficiency Analysis

SECURITY. Our GFPE scheme preserves only format-properties and hides all message-specific attributes. Thus, our scheme is as secure as the underlying integer-FPE scheme it employs. (This is a significant improvement over SGFPE, which was *less* secure than the underlying integer-FPE, from both theoretical and practical aspects.) We focus on the security of an implementation with FE1 or FE2 [2] as the underlying integer-FPE, because a rigourous security analysis exists only for FE1 and FE2, and not for FFX [3] or BPS [5]. The security of FE1, FE2 depends on their parameters. Specifically, they operate in rounds, in which a pseudo-random round function RF is applied to an intermediate value. Increasing the number of rounds reduces the adversarial advantage, and guarantee security against "stronger" adversaries (who can obtain more ciphertexts). By reducing the security of FE1 to the security of the round function RF, Bellare et al. [2] show it is SPI-secure, if sufficiently many rounds are used. Concretely, if RF is based on AES or SHA-256 then FE1 is "almost" as secure as AES or SHA-256, respectively. Regarding PRP-security, Bellare et al. [2] show that even a small number of rounds suffices to guarantee that the advantage of any efficient (i.e., poly-time) adversary is small (say, at most $2^{-80}$).

EFFICIENCY. Our scheme is most efficient when based on FFX, in which case it *may* incur cycle-walking, but the average cycle-length is at most 2. When provable security is required, or if one wishes to completely eliminate cycle walking, or to avoid using FFX (at least as long as it is not a NIST standard), our scheme can be based on FE1 or FE2 (see Figure 3, Section V for a comparison of the running time of our scheme when using FFX and FE1). In this case, our GFPE ranks strings in relation to their original format (without simplifying it), thus eliminating cycle-walking. Therefore, we can upper-bound the *actual* running time of encryption and decryption by $t = t_{\text{rank}} + t_{\text{intFPE}} + t_{\text{unrank}}$, where $t_X$ denotes the running time of algorithm X. This bound is *optimal* when using the RtE framework (whose advantages were already discussed). Our ranking and unranking algorithms are efficient, so the main efficiency bottleneck are integer-FPE encryption and decryption (which are "costly" operations). FE1 and FE2 are efficient when the size $N$ of the underlying format is not "too large", since these algorithms factor $N$. As the state-of-the-art factoring algorithms are super-polynomial, FE1 and FE2 should not be applied to very large formats. However, for many real-life formats (such as the address format, Example 4.2) factoring will render FE1 impractical. Therefore, to allow our GFPE to use FE1, we describe a method of *managing the size of general formats*, as discussed in the next section.[2]

## V. SUPPORTING LARGE FORMATS

Real-life formats, the main motivation for GFPE, are often so large that integer-FPE-based schemes become impractical, because all RtE-based FPE operations (namely ranking, un-ranking and integer-encryption or decryption) are computed in relation to the format size. As real-life format sizes are unlimited, if no upper-bound on permissable format sizes is enforced, any GFPE will soon become too inefficient to be used. (Even seemingly simple formats, e.g., the address formats of Example 4.2, are too large to be efficiently encrypted.) Despite being a major practical obstacle, known solutions ignore these issues. We remedy this by extending our framework to support large formats.

### A. Overview

At a high level, to guarantee efficiency for very large formats when using FE1, we need to insure that all operations are independent of the format size. This can only be achieved by restricting the format size, which causes security loss (see Section III-B), so minimizing this loss is a main goal. Since *any* restriction to sub-formats results in MP-insecure schemes (and consequently also PRP- and SPI-insecure), we focus on practical security, drawing conclusions from our study of SGFPE (Section III-B2). Concretely, by dividing a format *according to the sub-formats* from which it was constructed, and to *as few "pieces" as possible*, we try to hide message-specific attributes. Specifically, given a size bound maxS, larger formats are *split* recursively, where the compound format pre-processes (parses) the given plaintext, and delegates the splitting to its sub-formats. This raises the major question of how to split a format $\mathcal{F}$. Splitting $\mathcal{F} = \mathcal{F}_1 \cup ... \cup \mathcal{F}_k$ and applying RtE to each sub-format separately may incur insecurities (as in SGFPE, where splitting is based on message-specific properties). We propose a solution in which (using the parsability property) ranking is computed *recursively* in relation to sets of size at most maxS. The specific splitting strategy employed depends on the composition operations used to construct the compound format $\mathcal{F}$. The main underlying idea is to generalize the RtE framework to allow a message $m$ to be ranked to a *list* of ranks. Thus, given a format $\mathcal{F}$, and $m \in \mathcal{F}$, we can write $\mathcal{F}$ as the concatenation $\mathcal{F} = \mathcal{F}_1 \cdot ... \cdot \mathcal{F}_k$ where $|\mathcal{F}_i| < |\mathcal{F}|$, and parse $m$ accordingly: $m = m_1...m_k$. The encryption $c_i$ of each $m_i \in \mathcal{F}_i$ is then computed by RtE on $\mathcal{F}_i$, and the encryption of $m$ is $c := c_1...c_k$. Our splitting strategy has several advantages. First, ranking can be delegated to the underlying formats $\mathcal{F}_1, ..., \mathcal{F}_k$. Second, by choosing the sub-formats "correctly" (see Section V-B below), encryption hides many message-specific properties (since $m_i$ is encrypted to *any* ciphertext in $\mathcal{F}_i$). For example, the address format $\mathcal{F}_{\text{add}}$ of Example 4.2 can be split such that $\mathcal{F}_1 = \mathcal{F}_{\text{name}}$. Then the name in a given address is mapped to *any* valid name, and the message-specific properties discussed above remain entirely hidden. Splitting formats and encrypting ranks in relation to various sub-formats raises the following issue. The integer-FPE algorithms expect a size $M$ (determining the integer-domain $\mathcal{M}_{\text{int}} := \{0, 1, ..., M-1\}$) and an index $i \in \mathcal{M}_{\text{int}}$ to encrypt or decrypt. A message $m \in \mathcal{F}$ is represented by a list $r_1 \to r_2 \to ... \to r_n$ of ranks, where every $r_i$ was computed by the sub-format $\mathcal{F}_i$ in relation to $|\mathcal{F}_i|$. $|\mathcal{F}_1|, ..., |\mathcal{F}_1|$ *may be different*, and are possibly *unknown* to $\mathcal{F}$ (since ranking was performed by the sub-formats). Thus, we extend the ranking procedure to return not only a ranks list, but also a list of *sizes*, where entry $i$ contains the size of the integral-domain in relation to which the $i$'th rank was computed. Thus, integer-FPE encryption (and decryption) is performed on the size $M_i$ and the rank $r_i$. Due to technical reasons (which will become apparent in Section V-B below), the unranking method is given an "example" string with the same format as the string it should output, which for encryption (decryption) is the message to encrypt (the ciphertext to decrypt).

---
[2]We note that given a user-defined performance requirement, the appropriate value of maxS can be generated by the system by calculating the format size, and estimating FE1 runtime on it (since the runtime depends only on the format size). See our experiments in Section V for examples of format sizes.

## B. A GFPE For Large Formats

We describe the splitting operations for composition operations, and the generalized RtE framework. (Splitting algorithms for primitives, which are usually smaller and less likely to be split, are omitted due to space limitations.)

*1) Splitting Compound Formats:* In the following, maxS denote the upper bound on permissible format sizes.

UNION. Let $\mathcal{F} = \mathcal{F}_1 \cup ... \cup \mathcal{F}_k$.

- **Splitting.** $\mathcal{F}$ splits itself to sub-formats $\mathcal{F} = \mathcal{F}'_1 \cup ... \cup \mathcal{F}'_{k'}$ for some $k' \leq k$ as follows. Let $i_1 < i_2 < ... < i_{k'} \in \{1, 2, ..., k\}$ be indices such that for every $1 \leq i \leq k'$, either ($\sum_{l=i_j}^{i_{j+1}-1} |\mathcal{F}_j| \leq$ maxS and $\sum_{l=i_j}^{i_{j+1}} |\mathcal{F}_j| >$ maxS) or $i_{j+1} = i_j + 1$, then $\mathcal{F}'_j := \mathcal{F}_{i_j} \cup ... \cup \mathcal{F}_{i_{j+1}-1}$. (Intuitively, $\mathcal{F}_1, ..., \mathcal{F}_k$ are grouped into as few unions as possible such that every union does not exceed maxS, or (if that is not possible) the union consists of a single format $\mathcal{F}_i$.)

- **Ranking.** The rank of $m \in \mathcal{F}$ is computed by finding the index $1 \leq i \leq k'$ such that $m \in \mathcal{F}'_i$ and delegating ranking to $\mathcal{F}'_i$. If $|\mathcal{F}'_i| <$ maxS then ranking (of union formats, Section IV-C3) is performed directly by $\mathcal{F}'_i$, and ranking returns a ranks-list (sizes-list) containing a single rank $r$ (size $|\mathcal{F}'_i|$). Otherwise, $\mathcal{F}'_i = \mathcal{F}_j$ for some $1 \leq j \leq k$ such that $|\mathcal{F}_j| >$ maxS, and $\mathcal{F}_j$ first splits itself, before computing the rank.

- **Unranking.** The method receives a ranks-list $R$, a sizes-list $S$, and a message $f \in \mathcal{F}$, expected to have the same format as the plaintext $m$. (That is, if $m \in \mathcal{F}'_i$ then $\mathcal{F}$.unrank expects a string $f$ such that $f \in \mathcal{F}'_i$.) $f$ is used to determine an $1 \leq i \leq k'$ such that $m \in \mathcal{F}'_i$, and unranking is delegated to $\mathcal{F}'_i$.

CONCATENATION. We consider formats $\mathcal{F} = \mathcal{F}_1 \cdot d_1 \cdot \mathcal{F}_2 \cdot d_2 ... \cdot d_{k-1} \cdot \mathcal{F}_k$ (the algorithms for $\mathcal{F} = \mathcal{F}_1 \cdot ... \cdot \mathcal{F}_k$ are similar.)

- **Splitting.** $\mathcal{F}$ splits itself as $\mathcal{F} = \mathcal{F}'_1 \cdot ... \cdot \mathcal{F}'_{k'}$ for some $k' \leq k$, where $i_1 < i_2 < ... < i_{k'} \in \{1, 2, ..., k\}$ are such that for every $1 \leq i \leq k'$, either ($\prod_{l=i_j}^{i_{j+1}-1} |\mathcal{F}_j| \leq$ maxS and $\prod_{l=i_j}^{i_{j+1}} |\mathcal{F}_j| >$ maxS) or $i_{j+1} = i_j + 1$, then $\mathcal{F}'_j := \mathcal{F}_{i_j} \cdot d_{i_j} \cdot ... \cdot d_{i_{j+1}-2} \cdot \mathcal{F}_{i_{j+1}-1} \cdot d_{i_{j+1}-1}$.

- **Ranking.** To rank $m \in \mathcal{F}$ it is parsed to $m = m_1...m_{k'}$, where $\forall 1 \leq i \leq k', m_i \in \mathcal{F}'_i$ (this can be done since $m$ can be parsed according to $\mathcal{F}_1, ..., \mathcal{F}_k$, see Section IV-C3). Then, for every $1 \leq i \leq k'$, $r_i = \text{rank}(m_i)$ is computed in relation to $\mathcal{F}'_i$.

- **Unranking.** The unranking method receives a ranks-list $R$, a sizes-list $S$, and a message $f \in \mathcal{F}$, expected to have the same format of the plaintext $m$. The unranking method parses $f$ to $f = f_1...f_k$ where $f_i \in \mathcal{F}'_i$ for every $1 \leq i \leq k'$, and delegates the unranking to $\mathcal{F}'_1, ..., \mathcal{F}'_{k'}$ (the unranking of each $\mathcal{F}'_i$ is given $f_i$ and the corresponding entries in $R, S$).

RANGE. Let $\mathcal{F}$ be a range format defined by $\mathcal{F}', d, \min, \max$.

- **Splitting.** The format splits itself to sub-formats $\mathcal{F} = \mathcal{F}_1 \cup ... \cup \mathcal{F}_k$ as follows. Let $i_1 < i_2 < ... < i_k \in \{\min, \min + 1..., \max\}$ be such that for every $1 \leq i \leq k'$, either ($\sum_{l=i_j}^{i_{j+1}-1} |\mathcal{F}'|^l \leq$ maxS and $\sum_{l=i_j}^{i_{j+1}} |\mathcal{F}'|^l >$ maxS) or $i_{j+1} = i_j + 1$, then $\mathcal{F}'_j$ is defined as the range over $\mathcal{F}'$ with parameters $\mathcal{F}', d, i_j, i_{j+1} - 1$.

- **Ranking.** To rank an $m \in \mathcal{F}$, the index $1 \leq i \leq k'$ such that $m \in \mathcal{F}'_i$ is found, and the ranking of $m$ is delegated to $\mathcal{F}'_i$ (i.e., ranking finds the sub-range to which $m$ belongs, and $m$ is ranked in that sub-range).

- **Unranking.** The unranking method receives a ranks-list $R$, a sizes-list $S$, and a message $f \in \mathcal{F}$, expected to have the same format as the plaintext $m$. The method uses $f$ to determine the $1 \leq i \leq k'$ such that $m \in \mathcal{F}'_i$, and delegates the unranking operation to $\mathcal{F}'_i$.

*2) Encrypting Compound Formats:* We extend encryption and decryption to ranks-lists as follows. Given $\mathcal{F}$, and $m \in \mathcal{F}$, $\mathcal{F}.\text{rank}(m)$ returns a ranks-list $R = r_1 \to ... \to r_n$, and a sizes-list $S = s_1 \to ... \to s_n$. Then, integer-FPE encryption is invoked on every $r_i, s_i$ (encrypting $r_i$ in the domain $\{0, 1, ..., s_i - 1\}$), returning a list $R' = r'_1 \to ... \to r'_n$ of encrypted ranks. Finally, $\mathcal{F}.\text{unrank}(R', m)$ returns a list $c_1 \to ... \to c_n$ of unranked strings, and the encryption of $m$ is $c = c_1...c_n$. (Decryption is performed similarly, see the full version for further details.)

SECURITY. Since large formats are split into smaller sub-formats, the attacks of Section III-B apply to our scheme, which is unavoidable when bounding format sizes. However, by splitting formats according to their ingredients (rather than message-specific properties), our splitting strategy maintains the highest possible security level, and mostly hides message-specific properties that *do not* define the format.

*Example 5.1:* the format $\mathcal{F}_{\text{add}}$ (Example 4.2) can be split (say) to the sub-formats $\mathcal{F}'_1 = \mathcal{F}_{\text{name}} \cdot \mathcal{F}_{\text{num}} \cdot \mathcal{F}_{\text{space}}$, $\mathcal{F}'_2 = \mathcal{F}_{\text{dName}} \cdot \mathcal{F}_{\text{zip}} \cdot \mathcal{F}_{\text{space}}$, $\mathcal{F}'_3 = \mathcal{F}_{\text{state}}$. Thus, every substring is encrypted to *any* legal element in the corresponding sub-format (e.g., the encrypted name is *any* legal name, whereas SGFPE maintains *all* characteristics of the original name).

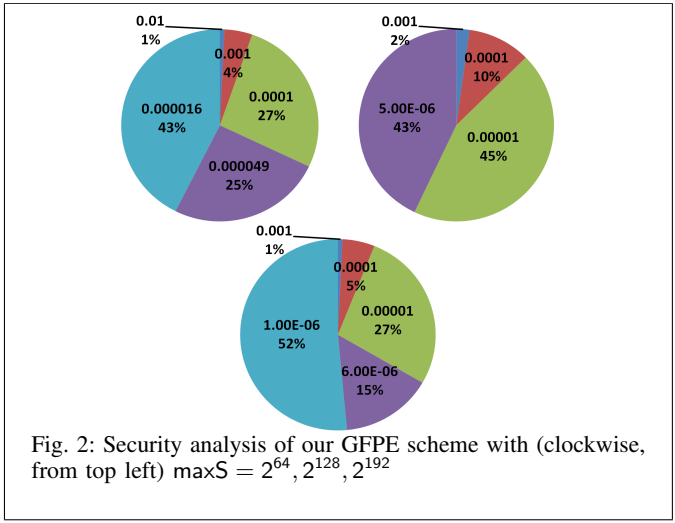

Fig. 2: Security analysis of our GFPE scheme with (clockwise, from top left) maxS $= 2^{64}, 2^{128}, 2^{192}$

The unavoidable disadvantage of splitting is that substring are encrypted independently. (Thus, in Example 5.1 two addresses with the same name and street number are encrypted to ciphertexts with the same (encrypted) name and street number.) Our scheme minimizes this security loss by splitting formats to the minimal number of "pieces", so if two addresses only share a name, their ciphertexts will have different names, since the name and street number are encrypted together (in SGFPE, both ciphertexts have the same name). maxS determines the concrete security level of our scheme. Concretely, if maxS $= 2^{128}$, then splitting reduces the message-space to size $2^{128}$, so no message-specific properties are revealed (except the message being in the smaller message-space), and the probability of recovering the message is at most $2^{-128}$, which can be ignored for all practical purposes. We stress, however, that maxS should be taken to be as large as possible (depending on the efficiency constraints). We tested our scheme on the FEC dataset (see Section III-B2) to determine the security loss. Our experiments show (Figure 2, top left) that even for maxS $= 2^{64}$ (as is the case in DES), only 4% of the donors can be identified with probability 0.001. (The probabilities are lower-bounds.) The top right (bottom) graph shows the if maxS $= 2^{128}$ (maxS $= 2^{192}$) then only 12% (6%) of the populace can be identified with probability at least 0.0001. (Additional graphs appear in the full version.) For maxS $= 2^{256}$ all message-specific attributes remain hidden, and the scheme is as secure as the underlying integer-FPE.

EFFICIENCY. We analyzed (Figure 3) the efficiency of our scheme on the FEC dataset with FFX and FE1, for various maxS values, and compared the performance with that of libFTE (Figure 4). Our experiments were performed on Intel-Core i5-3550 CPU 3.30GHz x4 processors with 7.6 Gib memory, using 64-bit Ubuntu 14.04 LTS. We used the FFX (FE1) implementation of libFTE [9] (Botan [7]). Encryption and decryption were performed on 100,000 messages. (The actual number of encrypted messages, as measured by #Messages, was larger due to splitting.) We measured the total ranking and unranking times (these are independent of the underlying integer-FPE scheme); running times of FFX and FE1 (on all messages); and total encryption times. Our experiments show that FFX achieves better performance; splitting significantly improves the running time of FE1 (and in some cases also of FFX) in this case (the corresponding format has size approximately $2^{856}$); but setting maxS $< 2^{256}$ has no efficiency gains. (We defer further experimental results to the full version.)

| MaxSize | #Messages | Rank | Unrank | FFX | | FE1 | |
|---|---|---|---|---|---|---|---|
| | | | | FFX Total | Overall | FE1 Total | Overall |
| - | 100000 | 26 | 126 | 98 | 275 | 1311 | 1486 |
| $2^{512}$ | 108238 | 27 | 80 | 84 | 213 | 638 | 746 |
| $2^{384}$ | 138504 | 26 | 66 | 107 | 225 | 446 | 540 |
| $2^{256}$ | 197319 | 26 | 63 | 131 | 253 | 276 | 367 |
| $2^{192}$ | 239902 | 26 | 63 | 124 | 252 | 299 | 390 |
| $2^{128}$ | 336471 | 26 | 65 | 164 | 317 | 403 | 496 |
| $2^{64}$ | 625143 | 24 | 68 | 318 | 504 | 726 | 820 |

Fig. 3: Running time (sec.) of our scheme with FFX and FE1

| Type | #Messages | Initialization | Rank | Unrank | FFX | Overall | Memory |
|---|---|---|---|---|---|---|---|
| libFTE (DFA) | 100000 | 0 | 1 | 8 | 110 | 121 | 113 MB |
| libFTE (NFA) | 100000 | 3 | 1697 | 15 | 100 | 1814 | 865 MB |
| Our Scheme | 108238 | - | 27 | 80 | 84 | 213 | 34 MB |

Fig. 4: Running time (sec.) of our scheme compared to libFTE

## VI. CONCLUSIONS

We propose a new general-format FPE scheme which (compared to existing methods) is more efficient (performance-wise), provides strong security guarantees, and, most importantly, is much more flexible in representing general formats. Our scheme is applicable and practical for solving real-life problems, e.g., delegating data and computation to the Cloud.